\documentclass[10pt,twoside]{Lowx2011}
\usepackage{epsf,amsmath}
\usepackage{graphicx}
\begin{document}

\title{String percolation and the first LHC data}
\author{I. Bautista $^{1,2}$ , J. Dias de Deus$^{2}$
and \underline{C. Pajares$^{1}$} \\ \\
%{\it for the Compostela Collaboration} \\ \\
\it $^{1}$Dpto. F\'\i sica de Part\'\i culas and\\
Instituto galego de F\'\i sica de Altas Enerx\'\i as\\
Universidade de Santiago de Compostela\\
Santiago de Compostela 15782\\
Galice, Spain\\
\it $^{2}$CENTRA, Departamento de F\'isica,\\
 IST, Av Rovisco Pais, \\
1049-001 Lisboa, Portugal.\\
E-mail: irais@fpaxp1.usc.es; jorge.dias.de.deus@ist.utl.pt;
pajares@fpaxp1.usc.es }

\maketitle

\begin{abstract}

\noindent The results of string percolation on multiplicities and elliptic
flow in AA and pp collisions are compared with LHC data showing a
good agreement. We discuss the rapidity long range correlations and
its relation to the height and longitudinal extension of the ridge
structure. Finally we show that the dependence of the shear
viscosity over entropy density ratio on the temperature, presents
a minimum close to the critical temperature remaining small in the
range of the RHIC and LHC energies.

\end{abstract}

\markboth{\large \sl \hspace*{0.25cm}\underline{C. Pajares} \& I. Bautista \& J. Dias de Deus
\hspace*{0.25cm} Low-$x$ Meeting 2011} {\large \sl \hspace*{0.25cm} String percolation and the first LHC data}

\section{String percolation}

The percolation of strings \cite{ref1}\cite{ref2}\cite{ref3} have described successfully
the basic facts, obtained at RHIC and LHC, of the Physics of QCD
matter at higher energy.

Strings are supposed to describe confined QCD interactions in an
effective way. They carry color charges at the ends and an
extended color field between the charges. They emit particles by
string breaking and pair creation. Projected in the impact
parameter they look like disks of radius $r_{0}\simeq 0.2$ fm and
two-dimensional percolation theory can be applied. As the energy
or the size of the projectile or target increases interaction
between strings occurs due to the overlapping of the strings and
the general result, due to the $SU(3)$ random summation of color
charges, is that there is a reduction in multiplicity, and an
increase in the string tension of formed clusters which means an
increase of $<p^{2}_{T}>$.

The relevant variable is the transverse string density $\eta_{t}$,
\begin{equation}
\eta_{t}\equiv \frac{\pi r^{2}_{0}}{S} N^{s}
\end{equation}
where $N^{s}$ is the number of strings and $S$ the overlapping
area. For $\eta_{t}$ larger than a critical value $\eta_{t}^{c}$ a
large cluster extends over the whole surface covering the fraction
$1-e^{-\eta_{t}}$ of the total area which at
$\eta_{t}=\eta^{c}_{t}$ is approximately $2/3$. For homogeneous
surface $\eta^{c}_{t}\simeq 1.2$ and for more realistic profiles
$\eta^{c}_{t}\simeq 1.5$ \cite{ref4}.

The basic formulae are, for particle density \cite{ref2}\cite{ref3}
\begin{equation}
\frac{dn}{dy}=F(\eta_{t})N^{s}\mu_{1}
\end{equation}
and for $<p^{2}_{T}>$
\begin{equation}
<p^{2}_{T}>=\frac{<p^{2}_{T}>_{1}}{F(\eta_{t})}
\end{equation}
where $F(\eta_{t})$ is the color reduction factor
\begin{equation}
F(\eta_{t})=\sqrt{\frac{1-e^{-\eta_{t}}}{\eta_{t}}}
\end{equation}
and $\mu_{1}$ and $<p^{2}_{T}>_{1}$ are the multiplicity and mean
$p^{2}_{T}$ produced by the fragmentation of a single string.

The multiplicity distribution can be obtained from the cluster size
distribution which approximately is a gamma function and the
multiplicity distribution of the cluster, which we assume
Poisson like. In this way, we obtain
the negative binomial distribution \cite{ref5}.
\begin{equation}
P(n,s)=\frac{\Gamma(n+k)}{\Gamma(n+1)\Gamma(k)}\frac{\gamma^{k}}{(1+\gamma)^{n+k}}
\mbox{ ,    } \gamma=\frac{k}{<n>},
\end{equation}
where $<n>$ is given by (5) and $k$ is identical to
\begin{equation}
\Re \equiv
\frac{<n(n-1)>-<n>^{2}}{<n>^{2}}=\frac{<N^{2}>-<N>^{2}}{<N>^{2}}=\frac{1}{k},
\end{equation}
where $N$ is the number of effective color sources and $\Re$ the normalized two particle correlation. Since, from (3) the size area of one effective cluster $\pi
r^{2}_{0}F(\eta_{t})$, and the area covered by strings is
$(1-e^{-\eta})\pi R^{2}$ $<N>$ is given by \cite{ref6}.
\begin{equation}
<N>=\frac{(1-e^{-\eta_{t}})R^{2}}{F(\eta_{t})r^{2}_{0}}=(1-e^{-\eta_{t}})^{1/2}\sqrt{\eta_{t}}(\frac{R}{r_{0}})^{2}
\end{equation}
Note that
\begin{equation}
<N>/N^{s}=F(\eta_{t}) \mbox{ , } <n>=<N>\mu_{1}
\end{equation}
We observe that in the low density limit, there is not overlapping
of strings and the particle density is essentially Poisson and we
have $k\rightarrow \infty$. In the large $\eta_{t}$ limit, the $N$
effective strings behave like a single string, with
$<N^{2}>-<N>^{2}\simeq <N>$ and therefore $k\rightarrow <N>
\rightarrow \infty$. At intermediate densities, $k$ has a minimum
close to the critical $\eta^{c}_{t}$, $k$ is given by \cite{ref6}.
\begin{equation}
k=\frac{<N>}{(1-e^{-\eta_{t}})^{3/2}}=\sqrt{\eta_{t}}(1-e^{-\eta_{t}})^{-1}(\frac{R}{r_{0}})^{2}
\end{equation}

In the glasma picture of color glass condensate also is obtained a
negative binomial distribution. In CGC the multiplicity is given
 by the number of color flux tube (strings),
$Q^{2}_{s}R^{2}_{A}$, times the number of gluons produced by one,
which is proportional to $1/\alpha_{s}(Q_{s})$. On the other hand,
k is the number of flux tubes, $Q^{2}_{s}R^{2}_{A}$, which in the
limit of high density coincides with (9) having the same $A$ and
$s$ dependencies.

Concerning the $p_{T}$ distributions, assuming a gaussian decay of
each cluster, whose width is given by (3) and taking into account
the gamma function distribution as the cluster size distribution
we obtain the following distribution \cite{ref6}[8-10]
\begin{equation}
\frac{dN}{dp^{2}_{T} dy}= \frac{dN}{dy} \frac{k'-1}{k'}
\frac{F(\eta_{t})}{<p^{2}_{T}>_{1}}(1+\frac{F(\eta_{t})p^{2}_{T}}{k'<p^{2}_{T}>_{1}})^{-k'}.
\end{equation}
This formula is not valid for high $p_{T}$ because we have assumed
a gaussian distribution for the decay of a cluster without any
power behavior corresponding to hard emissions. In (10) the $k'$
is a function of $\eta_{t}$ which has a qualitative similar
dependence that $k$ (may differ in the range of integration). At low $p_{T}$, (10) behaves like
$exp(\frac{-p^{2}_{T}}{<p^{2}_{T}>})$ and at moderate $p_{T}$ has
a power like behavior. The formula (10) is valid at all energies
and centralities including pp collisions, and gives a right
description of RHIC and LHC data, up to $p_{T}\simeq 5 $ GeV/c.

Note that at low energy density (10) behaves like
$exp(-p^{2}_{T}/<p^{2}_{T}>_{1})$. As the energy density
increases, $k'$ decreases and there is a departure of the
exponential behavior. However above critical point $k'$ increases
again and it is recover the exponential behavior.

\section{Multiplicity distributions}

The new LHC data has shown that the multiplicity at central
rapidity rise faster in central $Pb-Pb$ collisions than in pp
collisions. On the other hand, the data show the dependence in
the centrality of the multiplicity in AA is the same at LHC and
RHIC energies. These two facts are explained in our approach.

The apparently different behavior of the multiplicities is due to
energy conservation.In fact, in $Pb-Pb$ collisions the number of
strings grows like the number of collisions, $N_{A}^{4/3}$, and
the energy available only grows like A. Therefore, at not very
high energy, there will be strings that can not be formed because 
there is not energy available. We take into account this effect assuming that the
number of strings increases like $N_{A}^{1+\alpha(s)}$ where
$\alpha(s)$ is $0$ at low energy and goes to $1/3$ at very high
energy. In this way we obtain \cite{ref11}.
\begin{equation}
\frac{1}{N_{A}}\frac{dn}{y}|_{N_{A}N_{A}}=\frac{dn}{dy}|_{pp}[1+\frac{F(\eta^{t}_{N_{A}})}{F(\eta^{t}_{p})}(N^{\alpha(s)}_{A}-1)],
\end{equation}
with
\begin{equation}
\eta^{t}_{N_{A}}=\eta^{t}_{p}N^{\alpha}_{A}(\frac{A}{N^{2/3}_{A}})
\mbox{ ,    } \alpha=\frac{1}{3}(1-\frac{1}{1+ln(\sqrt{s/s_{0}}+1)})
\end{equation}
and
\begin{equation}
\frac{dn}{dy}|_{pp}=a(\frac{s}{m_{p}})^{\lambda/2}.
\end{equation}

In Fig.~1 we show our results for $\lambda=0.23$ together with the
data for AA and pp [12-14]. In Fig.~2, we show our results
concerning the centrality dependence for Pb-Pb at LHC energy and
Cu-Cu and Au-Au at RHIC energy.

\begin{figure}[ht]
\begin{center}
\includegraphics[width=.65\hsize]{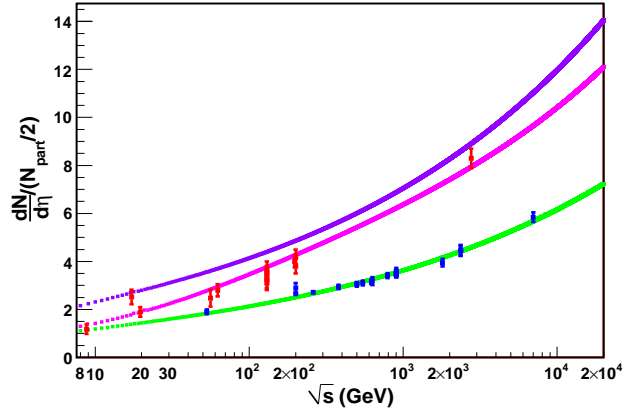}
\caption{ $dN_{ch}/d\eta$ form formula (11) at different
$\sqrt{s}$, lines in pink, and green correspond to Au-Au and p-p
collisions respectively. Line in purple shows the asymptotic
result for (11) with $F(\eta^{t})\rightarrow 1/\sqrt{\eta^{t}}$
and $\alpha \rightarrow 1/3$.}
\end{center}
\end{figure}

\begin{figure}[ht]
\begin{center}
\includegraphics[width=.65\hsize]{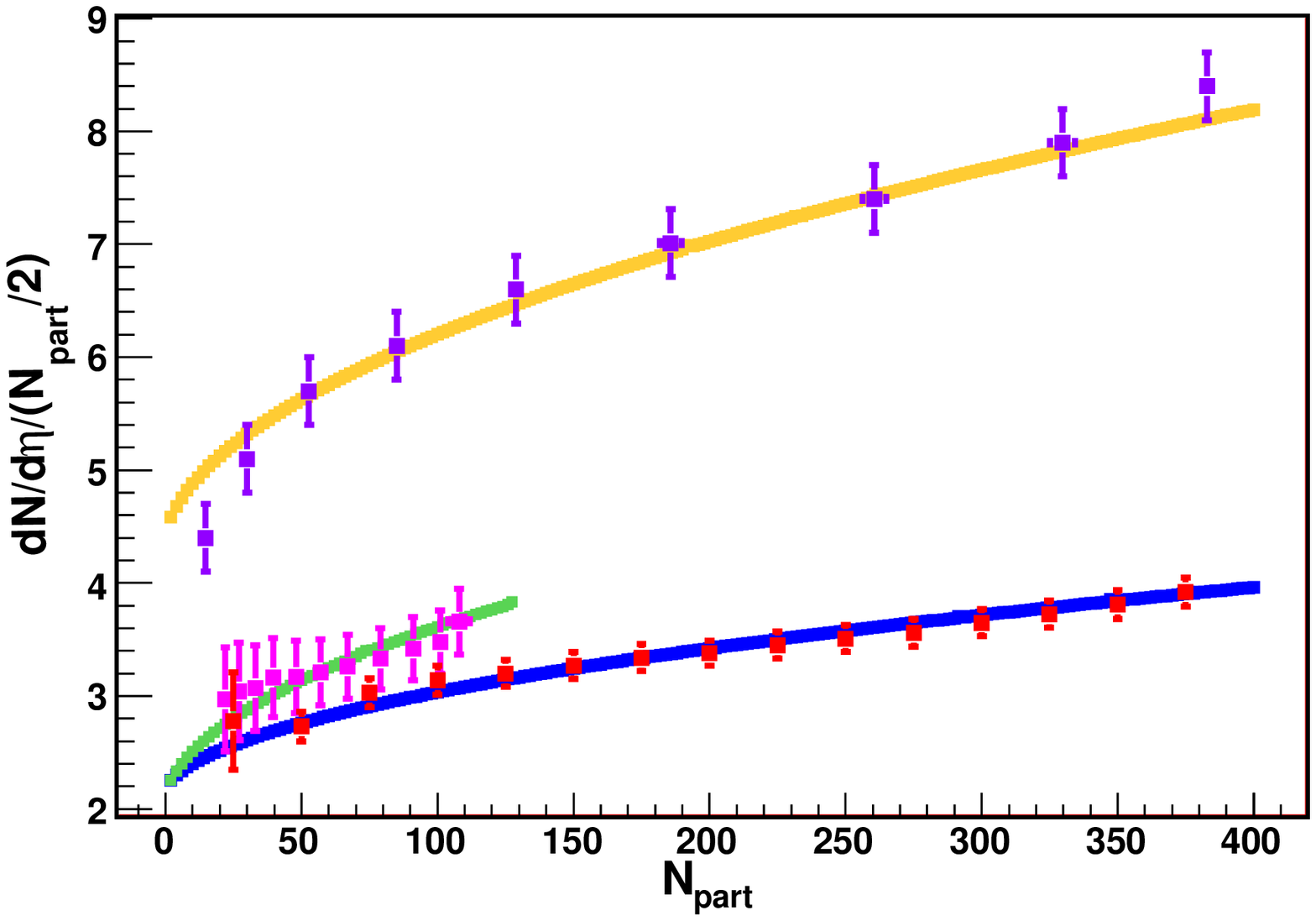}
\caption{ $dN_{ch}/d\eta$ from formula (11) at different
centralities at 200 GeV. Lines in blue and green are the
corresponding predictions for Au-Au and Cu-Cu. Points shown in red
and pink are the corresponding data for Au-Au and Cu-Cu from
reference at 200 GeV. Line in orange is the result of the formula
(11) at 2.76 TeV compared with ALICE data in purple.}
\end{center}
\end{figure}
Concerning the rapidity dependence, there is not limiting
fragmentation scaling in the percolation approach \cite{ref12} and the
evolution with energy is very different for different values of
the pseudorapidity $\eta$. This is clearly seen in Fig.~3 where is
shown $\frac{dn}{d\eta}|_{ch}$ for different $\eta$ values and
their energy dependence. It is seen that for central
pseudorapidity $\eta=0$ the energy dependence is weaker,
$s^{\lambda/2}$, than for high multiplicity $\eta\simeq 5$,
that is $s^{\lambda}$ \cite{ref15}.

\section{Long range rapidity correlations and the ridge structure}

The correlation introduced in formula (6) is relevant for the
discussion of rapidity long range correlations and the ridge
structure, first seen in Au-Au collisions and central Cu-Cu
collisions at RHIC as well in high multiplicity pp and Pb-Pb
collisions at LHC. In those collisions it was detected a
correlated broad peak of particles extended in rapidity and
localized in an azimuthal angle. The strength of the ridge
structure is proportional to $\Re \frac{dn}{dy}$.

The ridge structure seen in high multiplicity pp events \cite{ref16} was
predicted \cite{ref17}\cite{ref18} in the percolation approach, using equation
(12) that indicates that the string density for high multiplicity
events in pp at LHC is equivalent to the string density for Au-Au
peripheral collisions a $Cu-Cu$ central collisions at
$\sqrt{s}=200$ GeV.

In percolation, we have from (2) and (6) \cite{ref16}
\begin{equation}
\Re \frac{dn}{dy}=\frac{<N>}{k}=(1-e^{-\eta_{t}})^{3/2}
\end{equation}
As the density and/or energy increases the height of the near side
ridge structure increases slowly. Similar behavior is found in CGC
\cite{ref19} where
\begin{equation}
\Re \frac{dn}{dy}=\frac{1}{\alpha_{s}(Q_{s})}.
\end{equation}
The values of $\Re \frac{dn}{dy}$ are related to the value of the parameter $b$, 
which measures the rapidity long range correlations and it is defined by 
\begin{equation}
b=\frac{<n_{F}n_{B}>-<n_{F}><n_{B}>}{<n^{2}_{F}>-<n_{F}>^{2}}
\end{equation}
where $n_{F}$ and $n_{B}$ stand for the multiplicity in a forward and 
backward bins, separated by some rapidity gap to avoid short range correlations in the numerator of (17).

It is shown that \cite{ref20}\cite{ref21}
\begin{equation}
b=\frac{1}{1+\frac{k}{<N>}},
\end{equation}
and therefore 
\begin{equation}
b=\frac{1}{1+\frac{1}{\Re \frac{dn}{dy}}} = \frac{1}{1+(1-e^{-\eta_{t}})^{-3/2}}.
\end{equation}
At low energy density $b\rightarrow 0$, and at high density $b\rightarrow 1/2$. 
Similar behavior is obtained in CGC \cite{ref22}\cite{ref23}.

Notice that $k$ plays an important role in determining the long range rapidity correlations and also controls the inverse of the width of the KNO distribution $<n>P_{n}$ as a function of $n/<n>$.
As far as $k$ grows with energy density above the critical percolation value we predict that the width should decrease above this value. For minimum bias pp collisions the critical value is reached close to 
 $\sqrt{s}=14$ TeV. The data on pp shows in the measured range (up to 7 TeV) $k$ is decreasing \cite{ref24}.
 
 \section{Elliptic flow}
 In a precisely $b=0$, in AA or pp collisions the projected area in the impact parameter plane is a circle populated by disks approximately in a azimuthal uniform way. If $b \neq 0$ we have a projected almond. 
 If we imagine the projected almond to be obtained by a deformation of the circle, it is clear that the string density is larger along the smaller, $x$ axis, than the density along the $y$ axis. It is 
 intrinsic anisotropy that determines the existence of elliptic flow $v_{2}$. Also energy loss arguments support an sizable $v_{2}$. Notice that in percolation we have in the initial state interactions of the partons of the individual strings as a consequence of the color arrangement which is produced inside the formed cluster. The fragmentation of this large cluster produces a thermal distribution of particles \cite{ref30} and provide us the required early thermalization as far as the fragmentation time is around
  1 fm.
 
In order to compute $v_{2}$ we introduce the transverse azimuthal density [26-28].
\begin{equation}
\eta^{t}_{\phi}=\eta^{t}(\frac{R}{R_{\phi}})^{2}
\end{equation}
where 
\begin{equation}
R_{\phi}=R_{A}\frac{sin(\phi-\alpha)}{sin \phi}
\end{equation}
\begin{equation}
\alpha=sin^{-1}(\frac{b}{2 R_{A}}sin \phi)
\end{equation}
and 
\begin{equation}
\frac{\pi R^{2}_{A}}{4} \simeq \frac{1}{2} \int^{\pi/2}_{0} d\phi d R^{2}_{\phi}
\end{equation}
Introducing (20) into the transverse momentum distribution (10) and expanding the resulting 
distribution in powers of $(R^{2}_{\phi}-R^{2})$, retaining the first two terms, we obtain 
\begin{equation}
v_{2}(p^{2}_{T},y)=[\frac{2}{\pi}\int^{\pi/2}_{0} d\phi cos 2 \phi (\frac{R_{\phi}}{R})^{2}]
(\frac{e^{-\eta}
-F(\eta^{t})^{2}}{2 F(\eta^{t})})\frac{F(\eta^{t})p^{2}_{T}/<p^{2}_{T}>_{1}}{(1+F(\eta^{t})p^{2}_{T}/<p^{2}_{T}>_{1})}
\end{equation}
We observe that at low $p_{T}$ the dependence on $\eta^{t}$ is 
given by $(e^{-\eta_{t}}-F^{2}(\eta_{t}))/2F(\eta^{t})$ which
remain approximately constant for the values of $\eta^{t}$
corresponding to RHIC and LHC.
\begin{figure}[ht]
\begin{center}
    \begin{tabular}{cc}
      \resizebox{60mm}{!}{\includegraphics{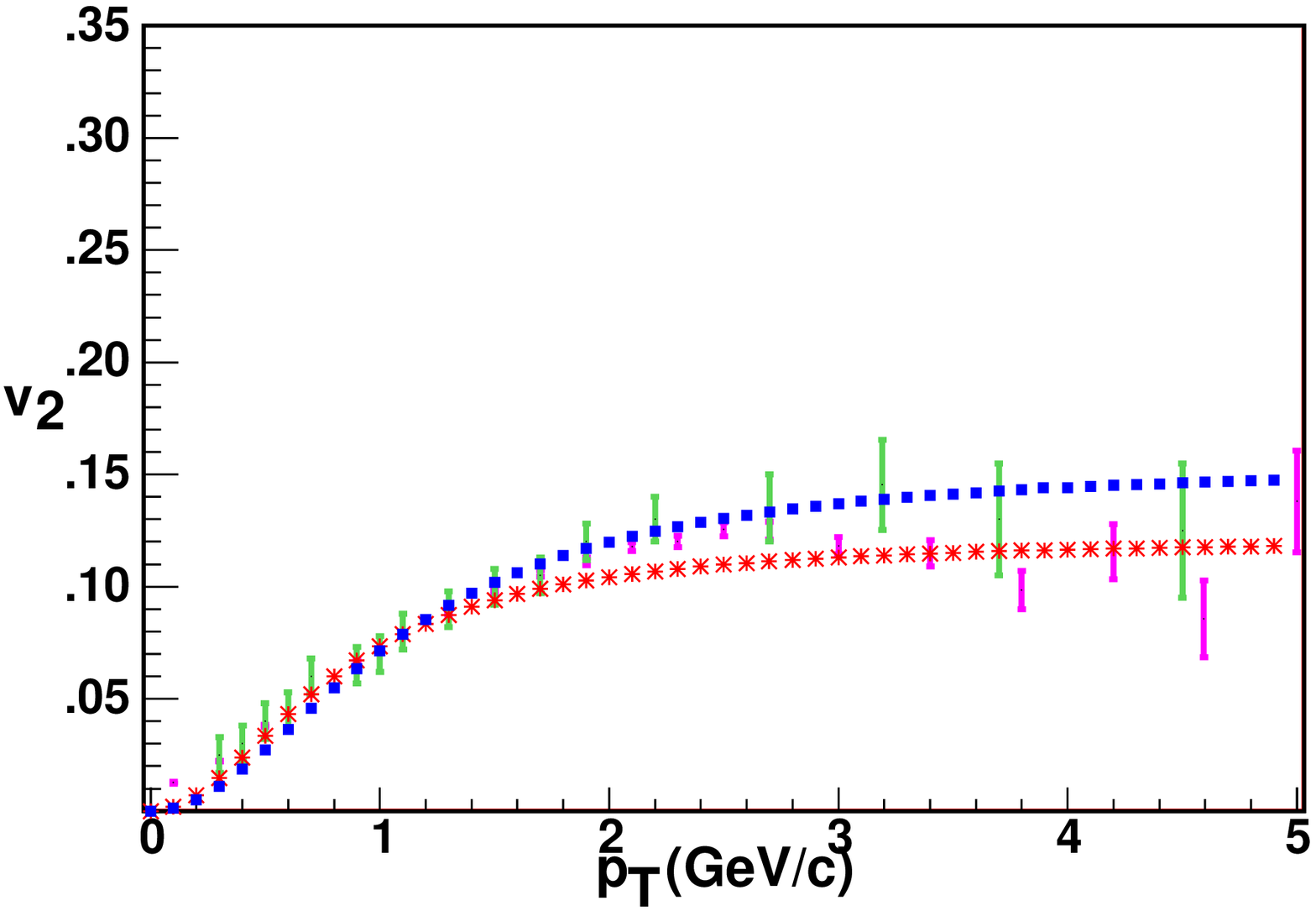}} &
      \resizebox{60mm}{!}{\includegraphics{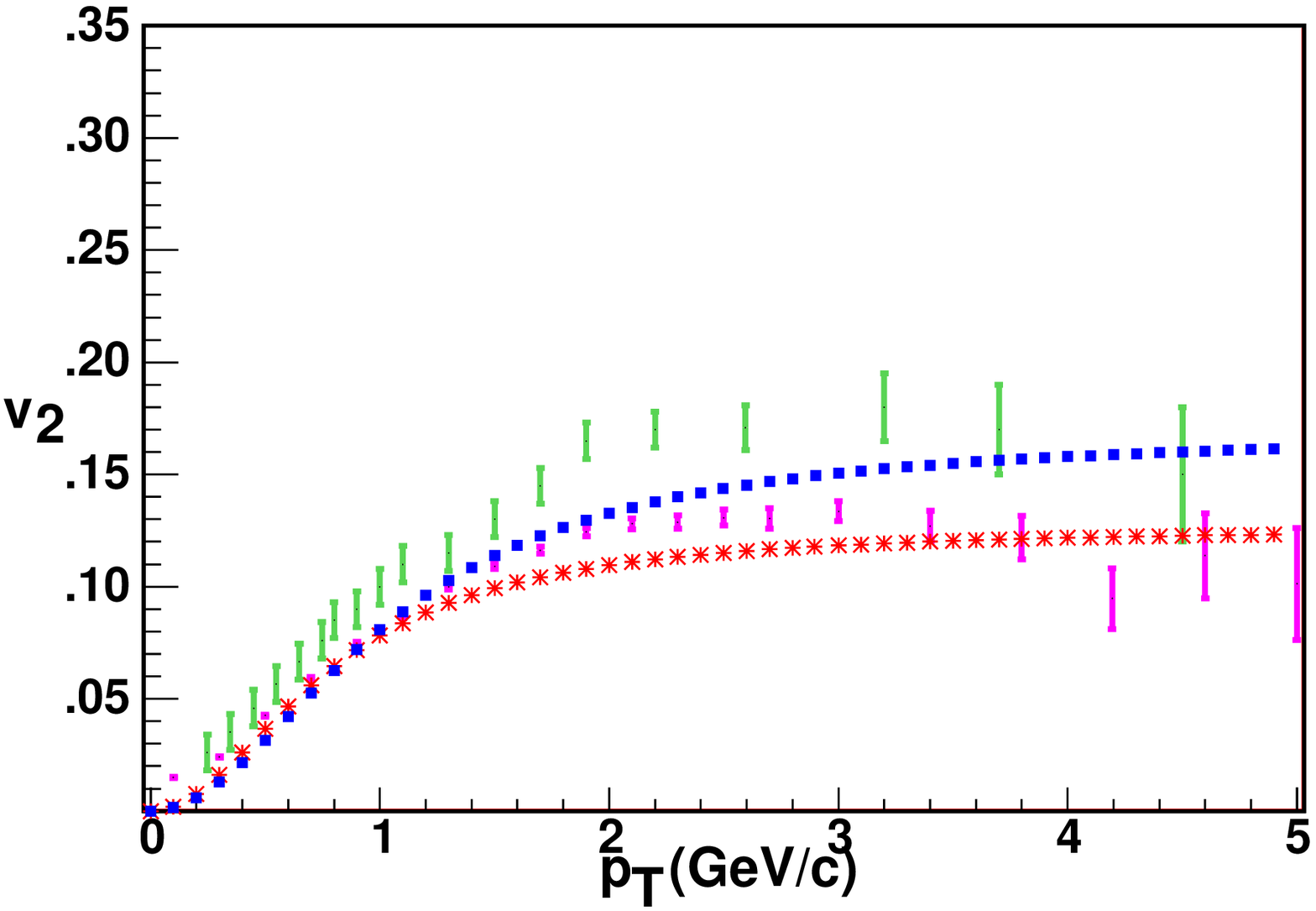}} \\
      \resizebox{60mm}{!}{\includegraphics{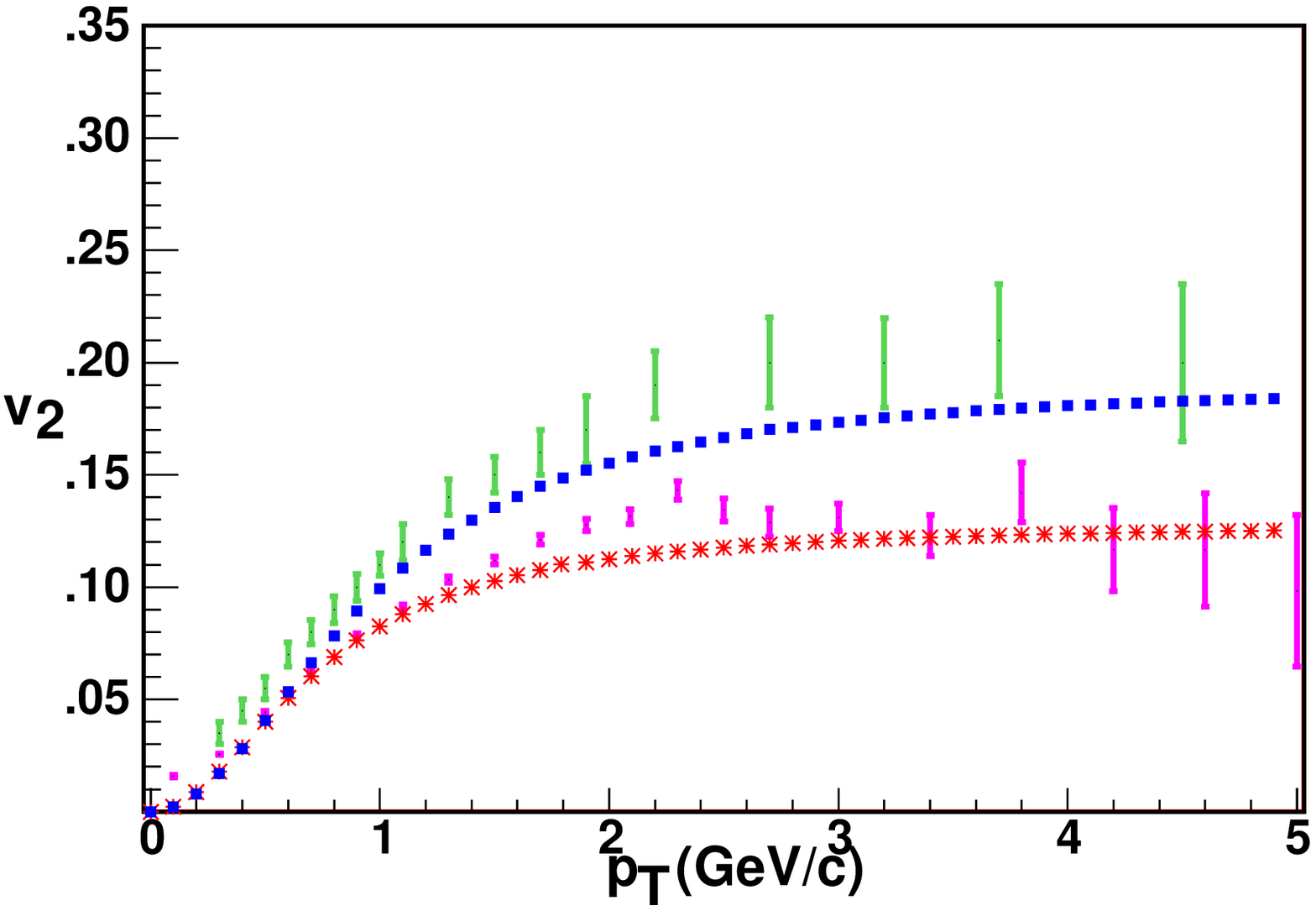}} &
      \resizebox{60mm}{!}{\includegraphics{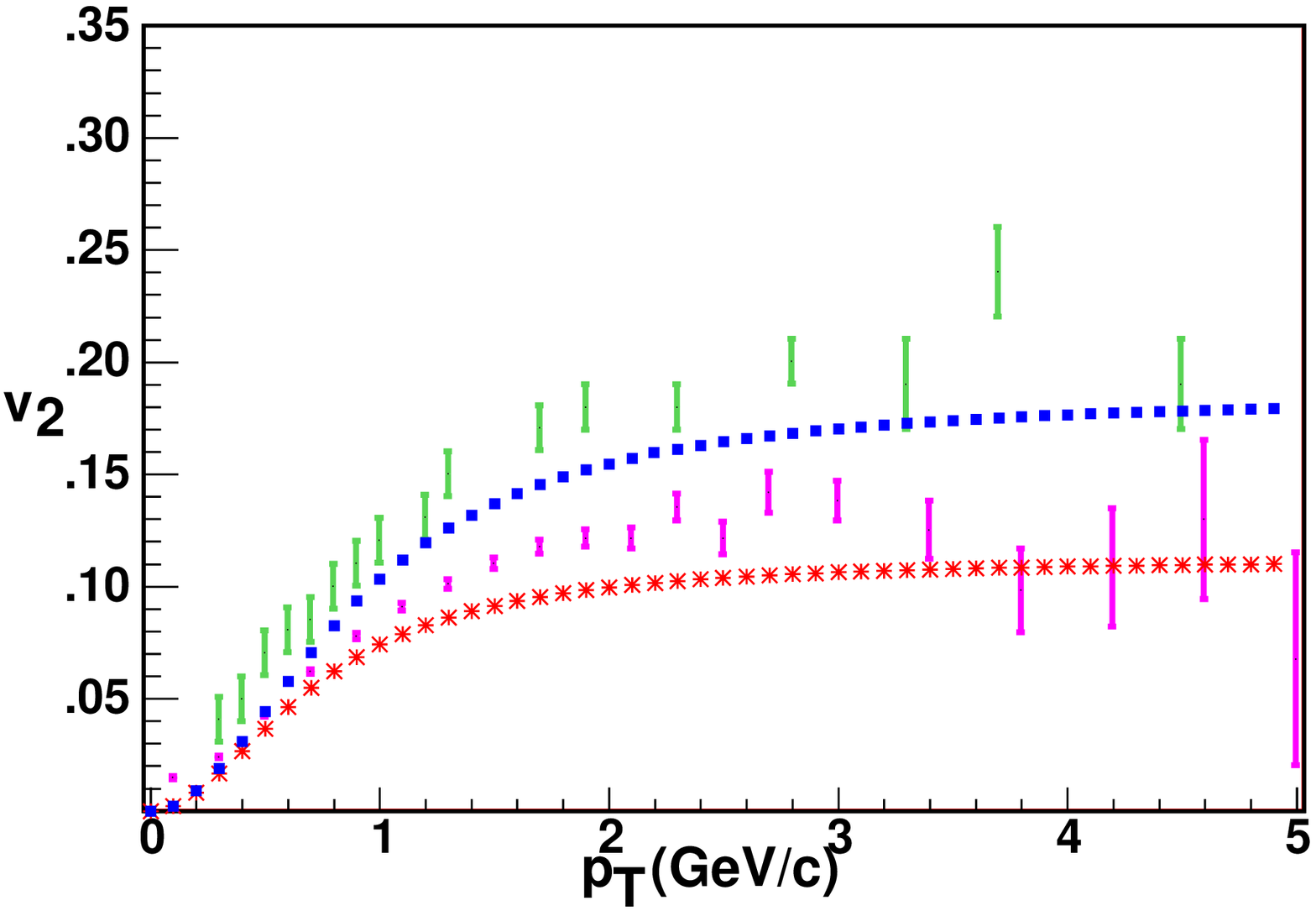}} \\
    \end{tabular}
    \caption{Color online stars in red and blue squares correspond to our
predictions for $\sqrt{s}=200$ GeV and $\sqrt{s}=2.76$ TeV
energies, and error-bars in green and pink are the respective data from RHIC and LHC for centralities  $10-20 \%$
, $20-30 \%$
, $30-40 \%$
, $40-50 \%$,
 figure caption order from top left to bottom right respectively.}
 \end{center}
 \end{figure}

\begin{figure}[ht]
\begin{center}
\includegraphics[width=.65\hsize]{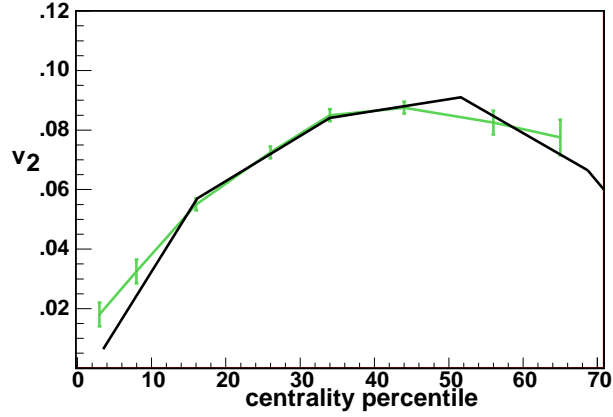}
\caption{Integrated $v_{2}$ at $\sqrt{s}=2.76$ TeV compared with
ALICE data.}
\end{center}
\end{figure}

In Fig.~3 we compare our results on the dependence of $v_{2}$ on
$p_{T}$ with RHIC and LHC data at different centralities \cite{ref12}-\cite{ref29}.
In Fig.~4 we compare the centrality dependence with ALICE
experimental data \cite{ref12}. In Fig.~5 we show our results for $\pi$,
$k$, and $p$ at $\sqrt{s}=2.76$ TeV. We obtain a very good
agreement with RHIC and LHC data, only our results on $p$ are
slightly different from data. Notice that our results are obtained
from the close universal formula (24) valid for all energies and
centralities. Apart from the parameter $<p^{2}_{T}>_{1}$ the only
input is the string density $\eta_{t}$ fixed from the
multiplicity.
\begin{figure}[ht]
\begin{center}
\includegraphics[width=.65\hsize]{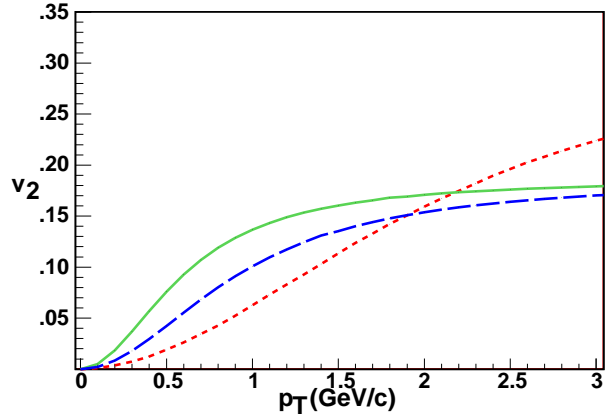}
\caption{Color online red doted line, green solid line and blue
dashed line are correspond to the proton, kaon, and pion
predictions for central $Pb-Pb$ collisions at $\sqrt{s}=2.76$
TeV.}
\end{center}
\end{figure}
\section{Shear viscosity entropy density ratio}

The shear viscosity to entropy density ratio is a measured of the
fluidity and the RHIC and LHC data show a low value, lower than
most of the known substances. In string percolation we obtain also
a low $\eta/s$ \cite{ref31}. In fact, in percolation the strong color
field inside the large cluster produces de-acceleration which can
be seen as a thermal temperature \cite{ref30}\cite{ref32} by means of Hawking-Unruh
effect. The temperature is given by
\begin{equation}
T(\eta^{t})=\sqrt{<p^{2}_{T}>_{1}/2F(\eta^{t})}
\end{equation}
On the other hand, from the relativistic kinetic theory $\eta/s$ is
given by
\begin{equation}
\frac{\eta}{s}=\frac{T\lambda}{5} \mbox {  } \lambda=
\frac{1}{n\sigma_{t r}}
\end{equation}
where $\lambda$ is the mean free path, n the number density and
$\sigma_{t r}$ the transport cross section.

From equation (7) we have
\begin{equation}
n=\frac{1-e^{-\eta^{t}}}{\pi r^{2}_{0}F(\eta^{t})L}
\end{equation}
and
\begin{equation}
\sigma_{t r}= S_{1}F(\eta^{t})=\frac{\pi r^{2}_{0}<p^{2}_{T}>_{1}}{2 T^{2}}
\end{equation}
From (26) and (27) we obtain
\begin{equation}
\frac{\eta}{s}=\frac{ L \sqrt{<p^{2}_{T}>_{1}} \eta^{1/4}_{t} }{5 \sqrt{2}(1-e^{-\eta_{t}})^{5/4}}
\end{equation}

\begin{figure}[ht]
\begin{center}
\includegraphics[width=.65\hsize, angle=270]{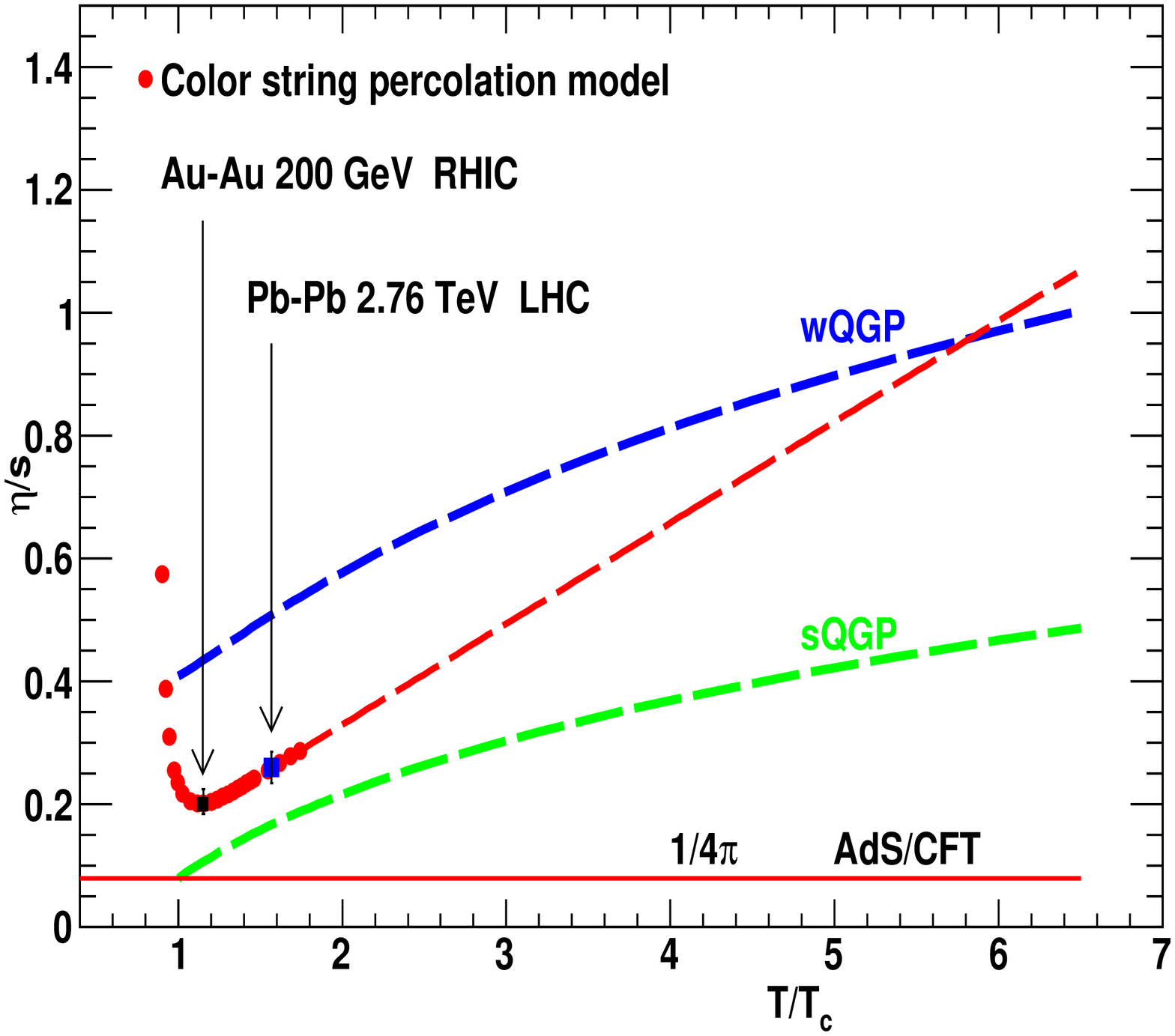}
\caption{}
\end{center}
\end{figure}
In Fig.~6 we show $\eta/s$ as a function of $T/Tc$. Close to the
critical temperature presents a minimum and remains small in the
RHIC and LHC range growing slowly.

\section*{Acknowledgements}

We thank J. G. Milhano, A. S. Hirsch, R. P. Scharenberg and B.
Srivastava who collaborates in part of the work reported here. We
thank the support of the FCT/Portugal project
PPCDT/FIS/5756682004, the project FPA2008-01177 of MICINN of
Spain, the Consolider project and the conselleria Eduacion da
Xunta de Galicia.

\end{document}